# Sabotage and Free Riding in Contests
# with a Group-Specific Public-Good/Bad Prize

*By* Kyung Hwan Baik and Dongwoo Lee[*]

February 2025

**Abstract**

We study contests in which two groups compete to win (or *not* to win) a group-specific public-good/bad prize. Each player in the groups can exert two types of effort: one to help her own group win the prize, and one to sabotage her own group's chances of winning it. The players in the groups choose their effort levels simultaneously and independently. We introduce a specific form of contest success function that determines each group's probability of winning the prize, taking into account players' sabotage activities. We show that two types of pure-strategy Nash equilibrium occur, depending on parameter values: one *without* sabotage activities and one *with* sabotage activities. In the first type, only the highest-valuation player in each group expends positive effort, whereas, in the second type, only the lowest-valuation player in each group expends positive effort.

[*]Baik: Department of Economics, Appalachian State University, Boone, NC 28608, USA, and Department of Economics, Sungkyunkwan University, Seoul 03063, South Korea (e-mail: khbaik@skku.edu); Lee (corresponding author): China Center for Behavioral Economics and Finance, Southwestern University of Finance and Economics, Chengdu, Sichuan 611130, China (e-mail: dwlee05@gmail.com). We are grateful to Chris Baik, Tara Bedi, Young-Chul Kim, and Freddie Papazyan for their helpful comments and suggestions. An earlier version of this paper was presented at the 99th Annual Conference of the Western Economic Association International, Seattle, WA, July 2024.

## 1. Introduction

Consider a situation in which a government (or private entity) tries to select a site for an airport and several regions (or cities) compete to be selected as the site for the airport. In the selected region, some residents benefit from hosting the airport because, for example, it creates jobs and stimulates local economies, while others are harmed by it due to noise, air pollution, increased traffic, etc. Naturally, residents in each region expend effort or make contributions to help their region be selected or to hinder it from being selected (or both). In this contest, we may well say that the airport is a group-specific prize because the residents in the selected region are affected by it (whether positively or negatively), whereas in the other regions that are not selected, residents are not affected. We may well say also that it is a public-good/bad prize because, in the selected region, it is a public good for some residents and a public bad for others.

Contests with a group-specific public-good/bad prize, like the motivating example above, are easily observed in the real world. Other examples include competition among locations to be selected as the site for a wind farm, a dam, a government institution, housing development, or a government-owned corporation.

The purpose of this paper is to study such contests.[1] Specifically, this paper studies two-group contests with a group-specific public-good/bad prize in which each individual in the groups can exert two types of effort: *constructive effort* to help her own group win the prize and *sabotage effort* to hinder her own group from winning the prize. We pose the following interesting questions. Who expends positive effort to help her own group win the prize? Who engages in sabotage activities? Are there any individuals who expend effort *both* to help her own group win the prize and to sabotage her own group's chances of winning it? How many individuals are there who expend positive effort? How much effort does each individual expend? How severe is the free-rider problem? What factors determine the effort levels expended by individuals?

To address these questions, this paper formally considers a model (or a game) in which two groups compete with each other to win (or *not* to win) a group-specific public-good/bad prize. Every player in the groups is risk-neutral, and her valuation for the prize is publicly known. In each group, there are at least one player with a positive valuation and at least one player with a negative valuation. Each player in the groups can exert two types of effort: constructive effort and sabotage effort. Each player has a constant marginal cost of increasing effort. All the players in the groups choose their effort levels simultaneously and independently.

This paper introduces a specific form of contest success function that determines each group's probability of winning the prize, taking into account players' sabotage activities. With this form, each group's probability of winning is determined by the two groups' *effective* effort levels, where each group's effective effort level equals the sum of constructive effort levels that the players in that group choose *minus* the sum of (effectiveness) adjusted sabotage effort levels of those players. Naturally, each group's effective effort level is positive, zero, or negative.

In Section 3, we first show that each player never expends *both* (positive) constructive effort and (positive) sabotage effort at a pure-strategy Nash equilibrium of the game. Specifically, each player with a positive valuation for the prize does not engage in sabotage activities, while each player with a negative valuation does not help her own group win the prize. Then, we show that every player in each group except the highest-valuation player [except the lowest-valuation player] expends zero effort at a pure-strategy Nash equilibrium.

In Section 4, we show that two types of pure-strategy Nash equilibrium occur, depending on parameter values: one *without* sabotage activities and one *with* sabotage activities. In the first type of equilibrium, only the highest-valuation player in each group is active – that is, expends positive effort – to *help* her own group win the prize. By contrast, in the second type, only the lowest-valuation player in each group is active to *hinder* her own group from winning the prize.

This paper is closely related to the literature on contests with a group-specific public-good prize – that is, the literature dealing with contests in which groups compete to win a prize, the prize is a public good for the players in the winning group, and the players in the losing groups are unaffected by the prize. Examples include Katz et al. (1990), Baik (1993, 2008), Baik et al. (2001), Epstein and Mealem (2009), Lee (2012), Kolmar and Rommeswinkel (2013),

Chowdhury et al. (2013), Topolyan (2014), Barbieri et al. (2014), Chowdhury and Topolyan (2016), Barbieri and Malueg (2016), Dasgupta and Neogi (2018), and Kobayashi (2019).

In this literature, the contest success function for a group is represented by a specific or general function of each group's (effective) effort level or by the selection rule of all-pay auctions, which is based on groups' (effective) effort levels. A group's (effective) effort level is assumed to equal the sum of effort levels that the players in the group expend, the minimum of effort levels that the players in the group expend (sometimes called the weakest link of the group), or the maximum of effort levels that the players in the group expend (sometimes called the best shot of the group). Most papers in this literature focus on examining the free-rider problem.

The most important and notable difference of this paper from all the previous papers in the literature is that, in this paper, the prize is assumed to be a public good for some players and a public bad for the others in the winning group, and the players are allowed to engage in sabotage activities against their own group, as well as to expend constructive effort to help their own group win the prize. By contrast, in all the previous papers, the prize is assumed to be a public good for all the players in the winning group, and each player in the groups is assumed to expend only constructive effort.

A strand of the literature on contests deals with sabotage activities. Konrad (2000) studies rent-seeking contests in which players can exert two types of effort: standard rent-seeking effort and sabotage effort. Amegashie and Runkel (2007) study an elimination contest in which players can sabotage potential rivals. Amegashie (2012) studies two-stage contests in which players expend sabotage effort in stage 1 and expend productive effort in stage 2. Amegashie (2013) reviews the literature on sabotage in contests. Chowdhury and Gürtler (2015) provide different perspectives on sabotage activities, and review the literature on contests with sabotage activities. Dogan et al. (2019) study team contests in which the members of a team exert not only their productive effort but also their sabotage effort directed at a particular member of the other team. Previous studies on contests involving sabotage activities have primarily focused on sabotage directed at opposing players or teams. By contrast, the current paper allows group members to have conflicting interests regarding the prize, which leads to within-group sabotage activities.

Baik (2023) has studied two-group contests with a group-specific public-good/bad prize. The model in the current paper differs from the one in Baik (2023), and the outcomes or predictions of the current paper differ from those of Baik (2023) in three respects. First, in the current paper, the players can hinder their own group from winning the prize by exerting sabotage effort directed toward their own group. By contrast, in Baik (2023), the players can do so by exerting constructive effort toward the other group's winning. In other words, they can decrease their own group's probability of winning by adding their constructive effort to the other group's effective effort level. Second, in the current paper, a group's effective effort level is assumed to equal the sum of constructive effort levels minus the sum of (effectiveness) adjusted sabotage effort levels of the players in that group. By contrast, in Baik (2023), it is assumed to equal the sum of constructive effort levels that the players in the group choose plus the sum of (effectiveness) adjusted constructive effort levels that the players in the other group choose to help this group win. Third, unlike Baik (2023), this paper introduces a new form of contest success function for a group, which takes into account sabotage effort toward the group.

The remainder of this paper proceeds as follows. Section 2 develops the model. In Section 3, we show several properties of a pure-strategy Nash equilibrium of the game, which also serve as preliminaries to Section 4. In Section 4, we obtain a pure-strategy Nash equilibrium of the game. Finally, Section 5 presents conclusions.

## 2. The model

There are two groups, 1 and 2, that compete to win (or *not* to win) a group-specific public-good/bad prize. The prize is a group-specific one because only the players in the winning group benefit from it or are harmed by it. The prize is a public-good/bad one because, in the winning group, it is a public good for some players and a public bad for the others. Each player in the groups can exert two types of effort: one to help her own group win the prize, and one to sabotage her own group's chances of winning it. The players in the groups choose their effort levels simultaneously and independently.

Group $i$, for $i = 1, 2$, consists of $n_i$ risk-neutral players, where $n_i \geq 2$. Let $N_i$ denote the set of players in group $i$: $N_i \equiv \{1, \ldots, n_i\}$. Let $v_{ik}$, for $k \in N_i$, denote the valuation for the prize of player $k$ in group $i$, where $v_{ik} > 0$ or $v_{ik} < 0$. The players' valuations for the prize are publicly known.

**Assumption 1.** ($a$) *We assume that* $v_{i1} > v_{i2} \geq \ldots \geq v_{in_i-1} > v_{in_i}$ *for* $i = 1, 2$. ($b$) *We assume that* $v_{i1}v_{in_i} < 0$ *for* $i = 1, 2$.

Part ($a$) of Assumption 1 assumes that there are only one highest-valuation player and only one lowest-valuation player in each group. This simplifies our analysis without affecting our main results. Part ($b$) assumes that, in each group, there are at least one player with a positive valuation and at least one player with a negative valuation − that is, the prize is a public good for some players and a public bad for the others.

Let $x_{ik}$, for $i = 1, 2$ and $k \in N_i$, denote the effort level that player $k$ in group $i$ chooses to help her own group win the prize, where $x_{ik} \geq 0$. Let $y_{ik}$ denote the effort level that player $k$ in group $i$ chooses to hinder her own group from winning the prize, where $y_{ik} \geq 0$. This is considered as her effort to sabotage her own group's chances of winning the prize. Both effort types are irreversible. Let $X_i \equiv \sum_{k=1}^{n_i} x_{ik}$ and $Y_i \equiv \sum_{k=1}^{n_i} y_{ik}$.

Let $p_i$, for $i = 1, 2$, denote the probability that group $i$ wins the prize. We assume the following contest success function for group $i$:

$$p_i = p_i(X_1 - \theta Y_1, X_2 - \theta Y_2),$$

where $\theta > 0$, $0 \leq p_i \leq 1$, and $p_1 + p_2 = 1$. The parameter $\theta$ reflects the relative effectiveness of effort expended in sabotage activities. We assume that the parameter $\theta$ is publicly known. We may say that each group's probability of winning depends on the two groups' *effective* effort levels, where group $i$'s effective effort level equals $X_i - \theta Y_i$.

More specifically, we assume the following contest success function for group 1:

$$p_1 = Z_1/(Z_1 + Z_2) \quad \text{for } Z_1 > 0 \text{ and } Z_2 \geq 0,$$
$$p_1 = (Z_1 + |Z_2|)/(Z_1 + |Z_2|) = 1 \quad \text{for } Z_1 \geq 0 \text{ and } Z_2 < 0,$$
$$p_1 = 0/(|Z_1| + Z_2) = 0 \quad \text{for } Z_1 \leq 0 \text{ and } Z_2 > 0, \tag{1}$$
$$p_1 = |Z_2|/(|Z_1| + |Z_2|) = 1 - |Z_1|/(|Z_1| + |Z_2|) \quad \text{for } Z_1 < 0 \text{ and } Z_2 \leq 0, \text{ and}$$
$$p_1 = 1/2 \quad \text{for } Z_1 = 0 \text{ and } Z_2 = 0,$$

where $Z_i \equiv X_i - \theta Y_i$ for $i = 1, 2$.[2] The contest success function for group 2 is specified by function (1) and the condition that $p_1 + p_2 = 1$.

Function (1) can be shortened as follows: $p_1 = \{\max(Z_1, 0) - \min(0, Z_2)\}/(|Z_1| + |Z_2|)$ for $|Z_1| + |Z_2| > 0$, and $p_1 = 1/2$ for $|Z_1| + |Z_2| = 0$. Function (1) assumes that, if group $i$'s effective effort level $Z_i$ is negative, then its absolute amount $|Z_i|$ is used to determine the groups' probabilities of winning the prize. Note that $|Z_2|/(|Z_1| + |Z_2|)$ can be interpreted as the probability that group 2 loses the prize. Function (1) has the properties similar to those of the simplest logit-form contest success function that is extensively used in the literature on the theory of contests.[3] For example, for $Z_1 > 0$ and $Z_2 > 0$, *ceteris paribus*, group $i$'s probability of winning is increasing in $Z_i$ at a decreasing rate: $\partial p_i/\partial Z_i > 0$ and $\partial^2 p_i/\partial Z_i^2 < 0$. It is decreasing in $Z_j$ at a decreasing rate: $\partial p_i/\partial Z_j < 0$ and $\partial^2 p_i/\partial Z_j^2 > 0$. Note that, for $Z_1 < 0$ and $Z_2 < 0$, we have that $\partial p_i/\partial Z_i > 0$ and $\partial^2 p_i/\partial Z_i^2 > 0$.

Let $\pi_{ik}$, for $i = 1, 2$ and $k \in N_i$, denote the expected payoff for player $k$ in group $i$. Then the payoff function for player $k$ in group $i$ is

$$\pi_{ik} = v_{ik} \, p_i(Z_1, Z_2) - x_{ik} - y_{ik}. \tag{2}$$

We consider the following noncooperative simultaneous-move game. At the start of the game, every player in both groups knows the values of $n_i$, $v_{ik}$, and $\theta$, for $i = 1$, 2 and $k \in N_i$. Next, the players in both groups choose their effort levels simultaneously and independently. Note that player $k$ in group $i$ chooses her effort levels, $x_{ik}$ and $y_{ik}$. Finally, the winning group is determined at the end of the game.

We assume that all of the above is common knowledge among the players. We employ Nash equilibrium as the solution concept.

## 3. No straddling, and free riding, in equilibrium

In this section, we first identify the pairs $(Z_1, Z_2)$ of the groups' effective effort levels which cannot occur at a pure-strategy Nash equilibrium of the game. Next, we show that, in equilibrium, each player never expends positive effort *both* to help her own group win the prize and to hinder her own group from winning the prize. Finally, we show that every player in each group except the highest-valuation player [except the lowest-valuation player] expends zero effort at a pure-strategy Nash equilibrium.

### 3.1. Pairs of the groups' effective effort levels which cannot occur in equilibrium

We begin by identifying the pairs $(Z_1, Z_2)$ of the groups' effective effort levels which cannot occur at a pure-strategy Nash equilibrium of the game.

**Lemma 1.** (*a*) *There exists no pure-strategy Nash equilibrium such that $Z_i > 0$ and $Z_j < 0$ for $i$, $j = 1$, 2 with $i \neq j$.* (*b*) *There exists no pure-strategy Nash equilibrium such that $Z_i = 0$ for some group $i = 1$, 2.*

**Proof.** (*a*) Consider a strategy profile, $(x_{11}, y_{11}, \ldots, x_{1n_1}, y_{1n_1}, x_{21}, y_{21}, \ldots, x_{2n_2}, y_{2n_2})$, such that $Z_i > 0$ and $Z_j < 0$ for $i$, $j = 1$, 2 with $i \neq j$. In this case, there exists a player in group $i$, say player $k$ for $k \in N_i$, such that $x_{ik} > 0$. According to (2), her payoff function is

$$\pi_{ik} \; = \; v_{ik}\, p_i(Z_1, Z_2) \; - \; x_{ik} \; - \; y_{ik}.$$

Then, using (1), it is straightforward to see that, *ceteris paribus*, her expected payoff increases when $x_{ik}$ decreases (by a small amount) – that is, she has an incentive to deviate from her effort level $x_{ik}$. This means that the strategy profile under consideration does not constitute a Nash equilibrium.

(*b*) Consider a strategy profile, $(x_{11}, y_{11}, \ldots, x_{1n_1}, y_{1n_1}, x_{21}, y_{21}, \ldots, x_{2n_2}, y_{2n_2})$, such that $Z_i = 0$ for some group $i = 1, 2$. First, suppose that $Z_j > 0$ for $j = 1, 2$ with $i \neq j$. In this case, there exists a player in group $j$, say player $k$ for $k \in N_j$, such that $x_{jk} > 0$. Then, using (1) and (2), it is straightforward to see that, *ceteris paribus*, her expected payoff increases when $x_{jk}$ decreases (by a small amount). This means that she has an incentive to deviate from her effort level $x_{jk}$, and thus the strategy profile under consideration does not constitute a Nash equilibrium.

Next, suppose that $Z_j = 0$ for $j = 1, 2$ with $i \neq j$. In this case, consider a player in group $j$, say player $k$ for $k \in N_j$, such that $v_{jk} > 0$ and $x_{jk} \geq 0$. Then, using (1) and (2), it is straightforward to see that, *ceteris paribus*, her expected payoff increases if she chooses $x_{jk} + \epsilon$, where $\epsilon$ is a positive infinitesimal. This implies that the strategy profile under consideration does not constitute a Nash equilibrium.

Finally, suppose that $Z_j < 0$ for $j = 1, 2$ with $i \neq j$. In this case, there exists a player in group $j$, say player $h$ for $h \in N_j$, such that $y_{jh} > 0$. Then, using (1) and (2), it is straightforward to see that, *ceteris paribus*, her expected payoff increases when $y_{jh}$ decreases (by a small amount). This implies that the strategy profile under consideration does not constitute a Nash equilibrium. $\qquad\qquad\square$

Due to Lemma 1, to obtain a pure-strategy Nash equilibrium of the game, it suffices to consider only the pairs $(Z_1, Z_2)$ of the groups' effective effort levels such that $Z_1 Z_2 > 0$. Hence, we will henceforth focus on such pairs of the groups' effective effort levels.

### 3.2. No straddling

Let $N_i^+$, for $i = 1, 2$, denote the set of players in group $i$ who have positive valuations for the prize. Let $N_i^-$ denote the set of players in group $i$ who have negative valuations. Part (*b*) of Assumption 1 implies that neither $N_i^+$ nor $N_i^-$ is empty.

**Lemma 2.** (*a*) *There exists no pure-strategy Nash equilibrium such that $y_{ik} > 0$ for some group $i = 1, 2$ and some player $k \in N_i^+$.* (*b*) *There exists no pure-strategy Nash equilibrium such that $x_{ih} > 0$ for some group $i = 1, 2$ and some player $h \in N_i^-$.*

**Proof.** (*a*) Consider a strategy profile, $(x_{11}, y_{11}, \ldots, x_{1n_1}, y_{1n_1}, x_{21}, y_{21}, \ldots, x_{2n_2}, y_{2n_2})$, such that $y_{ik} > 0$ for some group $i = 1, 2$ and some player $k \in N_i^+$. According to (2), the payoff function for player $k \in N_i^+$ is

$$\pi_{ik} = v_{ik}\, p_i(Z_1, Z_2) - x_{ik} - y_{ik}.$$

Then, using (1), it is straightforward to see that, *ceteris paribus*, her expected payoff increases when $y_{ik}$ decreases. This implies that the strategy profile under consideration does not constitute a Nash equilibrium.

(*b*) Consider a strategy profile, $(x_{11}, y_{11}, \ldots, x_{1n_1}, y_{1n_1}, x_{21}, y_{21}, \ldots, x_{2n_2}, y_{2n_2})$, such that $x_{ih} > 0$ for some group $i = 1, 2$ and some player $h \in N_i^-$. Then, using (1) and (2), it is straightforward to see that, *ceteris paribus*, the expected payoff for player $h \in N_i^-$ increases when $x_{ih}$ decreases. This implies that the strategy profile under consideration does not constitute a Nash equilibrium. □

Part (*a*) of Lemma 2 implies that, in equilibrium, each player with a positive valuation for the prize does not hinder her own group from winning the prize – in other words, she does not engage in sabotage activities. This makes an intuitive sense because, if her group wins the prize, she benefits from it; if her group loses it, she is unaffected by it.

Part (*b*) of Lemma 2 implies that, in equilibrium, each player with a negative valuation for the prize does not help her own group win the prize. This is natural because, if her group loses the prize, she suffers no harm; if her group wins it, she suffers a harm from it.

Due to Lemma 2, to obtain a pure-strategy Nash equilibrium of the game, it suffices to consider only the strategy profiles at which, for $i = 1, 2$, $y_{ik} = 0$ for $k \in N_i^+$ and $x_{ih} = 0$ for $h \in N_i^-$. Hence, we will henceforth focus on such strategy profiles.

### 3.3. Free riding

Based on Lemma 1, we consider only the following two cases: the case where $Z_i > 0$ for both $i = 1, 2$, and the case where $Z_i < 0$ for both $i = 1, 2$.

### 3.3.1. The case where $Z_1 > 0$ and $Z_2 > 0$

Player $k \in N_i^+$, for $i = 1, 2$, seeks to maximize her expected payoff

$$\pi_{ik} \ = \ v_{ik} \, p_i(Z_1, Z_2) \ - \ x_{ik} \ = \ v_{ik} \, Z_i/(Z_1 + Z_2) \ - \ x_{ik} \tag{3}$$

over her effort level $x_{ik}$, given effort levels of the other players in her own group and those of the players in the other group. Note that we set $y_{ik} = 0$ in (3) by taking Lemma 2 into account. Note also that player $k \in N_i^+$ has a positive valuation for the prize. Let $x_{ik}^b$ denote the best response of player $k \in N_i^+$ to a list of the other players' effort levels or, equivalently, her best response to a pair of $Z_{-ik}$ and $Z_j$, for $j = 1, 2$ with $i \neq j$, where $Z_{-ik} \equiv \sum_{z \neq k}(x_{iz} - \theta y_{iz})$. Then, it satisfies the first-order condition:

$$\partial \pi_{ik}/\partial x_{ik} \ = \ v_{ik} \, (\partial p_i/\partial Z_i) - 1 \ = \ v_{ik} \, Z_j/(Z_1 + Z_2)^2 - 1 \ = \ 0 \ \ \text{for } x_{ik}^b > 0. \tag{4}$$

It is straightforward to check that the payoff function, $\pi_{ik}$, is strictly concave in the effort level, $x_{ik}$. Hence, the second-order condition for maximizing $\pi_{ik}$ is satisfied, and $x_{ik}^b$ is unique.

Player $h \in N_i^-$, for $i = 1, 2$, seeks to maximize her expected payoff

$$\pi_{ih} = v_{ih} \, p_i(Z_1, Z_2) - y_{ih} = v_{ih} \, Z_i/(Z_1 + Z_2) - y_{ih} \qquad (5)$$

over her effort level $y_{ih}$, given $Z_{-ih}$ and $Z_j$, for $j = 1, 2$ with $i \neq j$. Note that we set $x_{ih} = 0$ in (5) by taking Lemma 2 into account. Note also that player $h \in N_i^-$ has a negative valuation for the prize. Using (5), we obtain

$$\partial \pi_{ih}/\partial y_{ih} = -\theta v_{ih} \, (\partial p_i/\partial Z_i) - 1 = -\theta v_{ih} \, Z_j/(Z_1 + Z_2)^2 - 1. \qquad (6)$$

Now, using (4) and (6), we obtain the following lemma. Let the superscript * denote the effort levels at a pure-strategy Nash equilibrium.

**Lemma 3.** (*a*) *There exists no pure-strategy Nash equilibrium such that $x_{ik} > 0$ for some group $i = 1, 2$ and some player $k \in N_i^+ \setminus \{1\}$. (b) There exists no pure-strategy Nash equilibrium such that $y_{ih} > 0$ for some group $i = 1, 2$ and some player $h \in N_i^-$.*

**Proof.** (*a*) Suppose on the contrary that there exists a pure-strategy Nash equilibrium, $(x_{11}^*, y_{11}^*, \ldots, x_{1n_1}^*, y_{1n_1}^*, x_{21}^*, y_{21}^*, \ldots, x_{2n_2}^*, y_{2n_2}^*)$, such that $x_{ik}^* > 0$ for some group $i = 1, 2$ and some player $k \in N_i^+ \setminus \{1\}$. Then, since $x_{ik}^*$ is her best response to the list of the other players' equilibrium effort levels, from (4) we have that $v_{ik}Z_j/(Z_1 + Z_2)^2 - 1 = 0$ at the Nash equilibrium. This, together with $v_{i1} > v_{ik} > 0$ from Assumption 1, yields that $\partial \pi_{i1}/\partial x_{i1} = v_{i1}Z_j/(Z_1 + Z_2)^2 - 1 > 0$ at the Nash equilibrium. This means that, *ceteris paribus*, the expected payoff for player 1 in group $i$ increases when $x_{i1}$ increases. This contradicts the assumption that $x_{i1}^*$ is her effort level at the Nash equilibrium.

(*b*) Suppose on the contrary that there exists a pure-strategy Nash equilibrium, $(x_{11}^*, y_{11}^*, \ldots, x_{1n_1}^*, y_{1n_1}^*, x_{21}^*, y_{21}^*, \ldots, x_{2n_2}^*, y_{2n_2}^*)$, such that $y_{ih}^* > 0$ for some group $i = 1, 2$ and some player $h \in N_i^-$. First, we have that $x_{i1}^* > 0$. This comes from part (*a*) of this lemma and the condition in this subsection that $Z_i > 0$. Then, since $x_{i1}^*$ is the best response of player 1 in group $i$ to the list of the other players' equilibrium effort levels, from (4) we have that $v_{i1}Z_j/(Z_1 + Z_2)^2 - 1 = 0$ at the Nash equilibrium. Second, from (6), we have that $\partial \pi_{ih} / \partial y_{ih} = -\theta v_{ih} \, Z_j / (Z_1 + Z_2)^2 - 1$. Now, from these two expressions, we have at the Nash equilibrium:[4]

$$\partial \pi_{ih} / \partial y_{ih} = -\theta v_{ih} \, Z_j / (Z_1 + Z_2)^2 - 1 \; < \; v_{i1} Z_j / (Z_1 + Z_2)^2 - 1 = 0 \quad \text{for} \; \theta |v_{ih}| < v_{i1}$$

and $\hspace{11cm}$ (7)

$$\partial \pi_{ih} / \partial y_{ih} = -\theta v_{ih} \, Z_j / (Z_1 + Z_2)^2 - 1 \; > \; v_{i1} Z_j / (Z_1 + Z_2)^2 - 1 = 0 \quad \text{for} \; \theta |v_{ih}| > v_{i1}.$$

The first case in (7) means that, *ceteris paribus*, the expected payoff for player $h \in N_i^-$ increases when $y_{ih}$ decreases. The second case in (7) means that, *ceteris paribus*, the expected payoff for player $h \in N_i^-$ increases when $y_{ih}$ increases. Both of these contradict the assumption that $y_{ih}^*$ is her effort level at the Nash equilibrium. $\hspace{4cm}$ $\square$

Part (*a*) of Lemma 3 can be explained as follows. Given $Z_j > 0$, group *i*'s optimal effective effort level $Z_i^b(k)$ for player $k \in N_i^+ \setminus \{1\}$ is smaller than that for player 1 in that group.[5] Hence, if $Z_i \geq Z_i^b(1)$ holds at a strategy profile, every player $k \in N_i^+ \setminus \{1\}$ has an incentive to decrease her effort level $x_{ik}$ (unless it is zero) since her marginal gross payoff, $v_{ik} Z_j / (Z_1 + Z_2)^2$, is less than her marginal cost, which is unity, at the strategy profile.

Part (*a*) of Lemma 3 implies that every player in $N_i^+$ except player 1 expends zero effort – namely, free rides on player 1's effort – at the pure-strategy Nash equilibrium with $Z_1^* > 0$ and $Z_2^* > 0$.

Part (*b*) of Lemma 3 can be explained as follows. Given $Z_{-ih} > 0$ and $Z_j > 0$, $\pi_{ih}$ is strictly convex in $y_{ih}$ over the interval $[0, Z_{-ih}/\theta)$ (see Section 4.1.1). Hence, in this interval, player $h \in N_i^-$ has an incentive to decrease her effort level $y_{ih}$ (unless it is zero) or increase it, depending on the values of $\theta$, $v_{ih}$, and $v_{i1}$ (see (7)).

Part (*b*) of Lemma 3 implies that every player in $N_i^-$ expends zero effort – and thus no sabotage occurs – at the pure-strategy Nash equilibrium with $Z_1^* > 0$ and $Z_2^* > 0$.

### 3.3.2. The case where $Z_1 < 0$ and $Z_2 < 0$

Player $h \in N_i^-$, for $i = 1, 2$, seeks to maximize her expected payoff

$$\pi_{ih} = v_{ih}\, p_i(Z_1, Z_2) - y_{ih} = v_{ih}\, \{1 - |Z_i|/(|Z_1| + |Z_2|)\} - y_{ih}$$
$$= v_{ih}\, \{1 - Z_i/(Z_1 + Z_2)\} - y_{ih}$$

over her effort level $y_{ih}$, given $Z_{-ih}$ and $Z_j$, for $j = 1,\, 2$ with $i \neq j$. Let $y_{ih}^b$ denote the best response of player $h \in N_i^-$ to a pair of $Z_{-ih}$ and $Z_j$. Then, it satisfies the first-order condition:

$$\partial \pi_{ih}/\partial y_{ih} = -\theta v_{ih}\, (\partial p_i/\partial Z_i) - 1 = \theta v_{ih}\, Z_j/(Z_1 + Z_2)^2 - 1 = 0 \quad \text{for } y_{ih}^b > 0. \tag{8}$$

It is straightforward to check that the payoff function, $\pi_{ih}$, is strictly concave in the effort level, $y_{ih}$. Hence, the second-order condition for maximizing $\pi_{ih}$ is satisfied, and $y_{ih}^b$ is unique.

Player $k \in N_i^+$, for $i = 1, 2$, seeks to maximize her expected payoff

$$\pi_{ik} = v_{ik}\, p_i(Z_1, Z_2) - x_{ik} = v_{ik}\{1 - |Z_i|/(|Z_1| + |Z_2|)\} - x_{ik}$$
$$= v_{ik}\{1 - Z_i/(Z_1 + Z_2)\} - x_{ik} \tag{9}$$

over her effort level $x_{ik}$, given $Z_{-ik}$ and $Z_j$, for $j = 1,\, 2$ with $i \neq j$. Using (9), we obtain

$$\partial \pi_{ik}/\partial x_{ik} = v_{ik}\, (\partial p_i/\partial Z_i) - 1 = -v_{ik}\, Z_j/(Z_1 + Z_2)^2 - 1. \tag{10}$$

Now, using (8) and (10), we obtain the following lemma.

**Lemma 4.** (*a*) *There exists no pure-strategy Nash equilibrium such that $y_{ih} > 0$ for some group $i = 1,\, 2$ and some player $h \in N_i^- \setminus \{n_i\}$. (b) There exists no pure-strategy Nash equilibrium such that $x_{ik} > 0$ for some group $i = 1,\, 2$ and some player $k \in N_i^+$.*

**Proof.** (*a*) Suppose on the contrary that there exists a pure-strategy Nash equilibrium, $(x_{11}^*, y_{11}^*, \ldots, x_{1n_1}^*, y_{1n_1}^*, x_{21}^*, y_{21}^*, \ldots, x_{2n_2}^*, y_{2n_2}^*)$, such that $y_{ih}^* > 0$ for some group $i = 1,\, 2$ and some player $h \in N_i^- \setminus \{n_i\}$. Then, since $y_{ih}^*$ is her best response to the list of the other players' equilibrium effort levels, from (8), we have that $\theta v_{ih} Z_j / (Z_1 + Z_2)^2 - 1 = 0$ at the Nash equilibrium. This, together with $v_{in_i} < v_{ih} < 0$ from Assumption 1, yields that $\partial \pi_{in_i} / \partial y_{in_i} = \theta v_{in_i} Z_j / (Z_1 + Z_2)^2 - 1 > 0$ at the Nash equilibrium. This means that, *ceteris paribus*, the expected payoff for player $n_i \in N_i^-$ increases when $y_{in_i}$ increases. This contradicts the assumption that $y_{in_i}^*$ is her effort level at the Nash equilibrium.

(*b*) Suppose on the contrary that there exists a pure-strategy Nash equilibrium, $(x_{11}^*, y_{11}^*, \ldots, x_{1n_1}^*, y_{1n_1}^*, x_{21}^*, y_{21}^*, \ldots, x_{2n_2}^*, y_{2n_2}^*)$, such that $x_{ik}^* > 0$ for some group $i = 1, 2$ and some player $k \in N_i^+$. First, we have that $y_{in_i}^* > 0$. This comes from part (*a*) of this lemma and the condition in this subsection that $Z_i < 0$. Then, since $y_{in_i}^*$ is the best response of player $n_i \in N_i^-$ to the list of the other players' equilibrium effort levels, from (8), we have that $\theta v_{in_i} Z_j / (Z_1 + Z_2)^2 - 1 = 0$ at the Nash equilibrium. Second, from (10), we have that $\partial \pi_{ik} / \partial x_{ik} = - v_{ik} Z_j / (Z_1 + Z_2)^2 - 1$. Now, from these two expressions, we have at the Nash equilibrium:[6]

$\partial \pi_{ik} / \partial x_{ik} = - v_{ik} Z_j / (Z_1 + Z_2)^2 - 1 \; < \; \theta v_{in_i} Z_j / (Z_1 + Z_2)^2 - 1 = 0 \;\; \text{for} \;\; v_{ik} < \theta |v_{in_i}|$

and (11)

$\partial \pi_{ik} / \partial x_{ik} = - v_{ik} Z_j / (Z_1 + Z_2)^2 - 1 \; > \; \theta v_{in_i} Z_j / (Z_1 + Z_2)^2 - 1 = 0 \;\; \text{for} \;\; v_{ik} > \theta |v_{in_i}|.$

The first case in (11) means that, *ceteris paribus*, the expected payoff for player $k \in N_i^+$ increases when $x_{ik}$ decreases. The second case in (11) means that, *ceteris paribus*, the expected payoff for player $k \in N_i^+$ increases when $x_{ik}$ increases. Both of these contradict the assumption that $x_{ik}^*$ is her effort level at the Nash equilibrium. □

The explanations of Lemma 4 are similar to those of Lemma 3, and therefore omitted. Part (*a*) of Lemma 4 implies that every player in $N_i^-$ except player $n_i$ expends zero effort − namely, free rides on player $n_i$'s sabotage effort − at the pure-strategy Nash equilibrium with $Z_1^* < 0$ and $Z_2^* < 0$. Part (*b*) of Lemma 4 implies that every player in $N_i^+$ expends zero effort − and thus there is only sabotage − at the pure-strategy Nash equilibrium with $Z_1^* < 0$ and $Z_2^* < 0$.

## 4. Two active players in equilibrium without or with sabotage activities

In this section, we obtain a pure-strategy Nash equilibrium of the game.[7]  There are two types of pure-strategy Nash equilibrium which do not occur simultaneously at any given set of parameter values: one *without* sabotage activities and one *with* sabotage activities.  In the first type of equilibrium, only the highest-valuation player in each group expends positive effort – while the rest expend zero effort – to *help* her own group win the prize.  By contrast, in the second type, only the lowest-valuation player in each group expends positive effort to *hinder* her own group from winning the prize.  We may say that, in the first type, the groups compete to win the prize; however, in the second type, they compete *not* to win the prize.

### 4.1. Preliminaries

As preliminary steps to obtain a pure-strategy Nash equilibrium, we first look at a best response of player $h$ in $N_i^-$, for $i = 1, 2$, to a pair of $Z_{-ih}$ and $Z_j$, for $j = 1, 2$ with $i \neq j$, in the case where $Z_{-ih} > 0$ and $Z_j > 0$, and then look at a best response of player $k$ in $N_i^+$ to a pair of $Z_{-ik}$ and $Z_j$, in the case where $Z_{-ik} < 0$ and $Z_j < 0$.

### 4.1.1. The case where $Z_{-ih} > 0$ and $Z_j > 0$

In this case, the payoff function for player $h$ in $N_i^-$ is

$$\pi_{ih} = v_{ih} Z_i / (Z_1 + Z_2) - y_{ih} \qquad \text{for } 0 \leq y_{ih} < Z_{-ih}/\theta$$
$$- y_{ih} \qquad\qquad\qquad \text{for } y_{ih} \geq Z_{-ih}/\theta.$$

For $0 \leq y_{ih} < Z_{-ih}/\theta$, we obtain

$$\partial^2 \pi_{ih} / \partial y_{ih}^2 = -2\theta^2 v_{ih} Z_j / (Z_1 + Z_2)^3 > 0,$$

which means that $\pi_{ih}$ is *strictly convex* in $y_{ih}$.

Accordingly, the expected payoff of player $h$ in $N_i^-$ is maximized at $y_{ih} = 0$ or $y_{ih} = Z_{-ih}/\theta$ (or both).  Since $y_{ih} = 0$ yields $\pi_{ih} = v_{ih} Z_{-ih}/(Z_{-ih} + Z_j)$ and $y_{ih} = Z_{-ih}/\theta$ yields $\pi_{ih} = -Z_{-ih}/\theta$, her best response is 0 if $\theta|v_{ih}| \leq Z_{-ih} + Z_j$, which comes from the condition that $v_{ih}Z_{-ih}/(Z_{-ih} + Z_j) \geq -Z_{-ih}/\theta$; and it is $Z_{-ih}/\theta$ if $\theta|v_{ih}| \geq Z_{-ih} + Z_j$.

Lemma 5 summarizes this result.

**Lemma 5.** *Suppose that $Z_{-ih} > 0$ and $Z_j > 0$. Then, a best response of player $h \in N_i^-$ to a pair of $Z_{-ih}$ and $Z_j$ is 0 if $\theta|v_{ih}| \leq Z_{-ih} + Z_j$, and it is $Z_{-ih}/\theta$ if $\theta|v_{ih}| \geq Z_{-ih} + Z_j$.*

*4.1.2. The case where $Z_{-ik} < 0$ and $Z_j < 0$*

In this case, the payoff function for player $k$ in $N_i^+$ is

$$\pi_{ik} = v_{ik}\{1 - Z_i/(Z_1 + Z_2)\} - x_{ik} \qquad \text{for } 0 \leq x_{ik} < |Z_{-ik}|$$
$$v_{ik} - x_{ik} \qquad \text{for } x_{ik} \geq |Z_{-ik}|.$$

For $0 \leq x_{ik} < |Z_{-ik}|$, we obtain

$$\partial^2\pi_{ik}/\partial x_{ik}^2 = 2v_{ik} Z_j/(Z_1 + Z_2)^3 > 0,$$

which means that $\pi_{ik}$ is *strictly convex* in $x_{ik}$.

Accordingly, the expected payoff of player $k$ in $N_i^+$ is maximized at $x_{ik} = 0$ or $x_{ik} = |Z_{-ik}|$ (or both). Since $x_{ik} = 0$ yields $\pi_{ik} = v_{ik}\{1 - Z_{-ik}/(Z_{-ik} + Z_j)\}$ and $x_{ik} = |Z_{-ik}|$ yields $\pi_{ik} = v_{ik} - |Z_{-ik}|$, her best response is 0 if $v_{ik} \leq |Z_{-ik} + Z_j|$, which comes from the condition that $v_{ik}\{1 - Z_{-ik}/(Z_{-ik} + Z_j)\} \geq v_{ik} - |Z_{-ik}|$; and it is $|Z_{-ik}|$ if $v_{ik} \geq |Z_{-ik} + Z_j|$.

Lemma 6 summarizes this result.

**Lemma 6.** *Suppose that $Z_{-ik} < 0$ and $Z_j < 0$. Then, a best response of player $k \in N_i^+$ to a pair of $Z_{-ik}$ and $Z_j$ is 0 if $v_{ik} \leq |Z_{-ik} + Z_j|$, and it is $|Z_{-ik}|$ if $v_{ik} \geq |Z_{-ik} + Z_j|$.*

### 4.2. Pure-strategy Nash equilibrium of the game

We are now ready to obtain the pure-strategy Nash equilibrium of the game. Note that, at the Nash equilibrium, each player's pair of effort levels, $(x^*_{ik}, y^*_{ik})$ for $i = 1, 2$ and $k \in N_i$, is a best response to the other players' pairs of effort levels.

### 4.2.1. The case where $Z^*_1 > 0$ and $Z^*_2 > 0$

Using Lemmas 2, 3, and 5, we obtain Proposition 1.

**Proposition 1.** *Suppose that $\max\{\theta|v_{1n_1}|, \; \theta|v_{2n_2}|\} \leq v_{11}v_{21}/(v_{11} + v_{21})$ or, equivalently, $\theta \leq v_{11}v_{21}/[(v_{11} + v_{21}) \cdot \max\{|v_{1n_1}|, \; |v_{2n_2}|\}]$. Then, there exists a unique pure-strategy Nash equilibrium, $(x^*_{11}, y^*_{11}, \ldots, x^*_{1n_1}, y^*_{1n_1}, x^*_{21}, y^*_{21}, \ldots, x^*_{2n_2}, y^*_{2n_2})$, such that, for $i = 1, 2$, $x^*_{11} = v^2_{11}v_{21}/(v_{11} + v_{21})^2$, $x^*_{21} = v_{11}v^2_{21}/(v_{11} + v_{21})^2$, $x^*_{ik} = 0$ for $k \in N_i \setminus \{1\}$, and $y^*_{ik} = 0$ for $k \in N_i$. In this case, we have: $X^*_i = x^*_{i1}$, $Y^*_i = 0$, and thus $Z^*_i = x^*_{i1} > 0$.*

**Proof**. First, it is straightforward to check that $(x^*_{i1}, y^*_{i1})$, for $i = 1, 2$, is the best response of player 1 in group $i$ to the other players' equilibrium pairs of effort levels. Indeed, it is straightforward to check that $x^*_{i1}$ is her best response to the pair of $Z^*_{-i1}$ and $Z^*_j$, for $j = 1, 2$ with $i \neq j$, and $y^*_{i1} = 0$ is her best response to the pair of $Z^*_{-i1}$ and $Z^*_j$ (see the proof of part (*a*) of Lemma 2).

Next, it is also straightforward to check that $(x^*_{ik}, y^*_{ik})$, for $i = 1, 2$ and $k \in N^+_i \setminus \{1\}$, is the best response of player $k$ to the other players' equilibrium pairs of effort levels (see the proof of part (*a*) of Lemma 3 and that of Lemma 2).

Finally, we need to show that $(x^*_{ih}, y^*_{ih})$, for $i = 1, 2$ and $h \in N^-_i$, is a best response of player $h$ to the other players' equilibrium pairs of effort levels. It is immediate from the proof of part (*b*) of Lemma 2 that $x^*_{ih} = 0$ is her best response to the pair of $Z^*_{-ih}$ and $Z^*_j$.

We now show that $y^*_{ih}$ is her best response to the pair of $Z^*_{-ih}$ and $Z^*_j$. Using Lemma 5, we know that $y^*_{ih} = 0$ is her best response to the pair of $Z^*_{-ih}$ and $Z^*_j$ if $\theta|v_{ih}| \leq Z^*_{-ih} + Z^*_j$ $= x_{11}^* + x_{21}^* = v_{11}v_{21}/(v_{11} + v_{21})$.[8] This leads, under Assumption 1, to the fact that, for *every* player $h \in N_i^-$, $y_{ih}^*$ is her best response to the pair of $Z_{-ih}^*$ and $Z_j^*$ if $\theta|v_{in_i}| \leq v_{11}v_{21}/(v_{11} + v_{21})$.

Therefore, for *both* $i = 1, 2$ and *every* player $h \in N_i^-$, $y_{ih}^*$ is her best response to the pair of $Z_{-ih}^*$ and $Z_j^*$ under the given condition that $\max\{\theta|v_{1n_1}|, \theta|v_{2n_2}|\} \leq v_{11}v_{21}/(v_{11} + v_{21})$. □

Note that $v_{11}v_{21}/(v_{11} + v_{21}) < v_{i1}$ for $i = 1, 2$. Let $w = \max\{\theta|v_{1n_1}|, \theta|v_{2n_2}|\}$. Figure 1 illustrates, with the shaded area, the values of $v_{11}$ and $v_{21}$ which satisfy the condition that $w \leq v_{11}v_{21}/(v_{11} + v_{21})$, given a value of $w$. Hence, the pure-strategy Nash equilibrium identified in Proposition 1 occurs at the values of $v_{11}$ and $v_{21}$ located in the shaded area of Figure 1.

If $w$ increases, then the curve representing the equation $v_{11}v_{21}/(v_{11} + v_{21}) = w$ in Figure 1 shifts outward, reducing the shaded area. For example, if the parameter $\theta$ increases, *ceteris paribus*, then the set of $(v_{11}, v_{21})$ shrinks at which the pure-strategy Nash equilibrium identified in Proposition 1 occurs.[9] This makes intuitive sense. An increase in the parameter $\theta$ makes effort expended in sabotage activities more effective. As a result, player $n_i$'s best response to a pair of $Z_{-in_i}$ and $Z_j$ may change from 0 to $Z_{-in_i}/\theta$, depending on the given valuation profile (see Lemma 5).

Proposition 1 says that, if $\max\{\theta|v_{1n_1}|, \theta|v_{2n_2}|\} \leq v_{11}v_{21}/(v_{11} + v_{21})$, then there exists a pure-strategy Nash equilibrium with the following properties. First, there are only two active players, one from each group. The active player in each group is player 1 in that group – that is, the highest-valuation player in the group. Second, every player in $N_i^+$ except player 1 free rides on player 1's effort because she expects player 1's effort level to be large enough from her perspective (see Lemma 3). Third, every player in $N_i^-$ gives up on sabotaging her own group because her "adjusted" valuation $\theta|v_{ih}|$ for the prize is far smaller than the valuation of player 1, the only active player, in her own group, and also because it is far smaller than the valuation of player 1, the only active player, in the other group. To put it simply, she gives up on sabotaging because she cannot defeat player 1's desire, in her own group, to win the prize, and also because player 1's desire, in the other group, to win the prize is strong enough from her perspective.

At the pure-strategy Nash equilibrium identified in Proposition 1, the two active players (or the two groups) compete to win the prize rather than to lose it. We may simply say that the effort level $x_{11}^*$ of player 1 in group 1 is the best response to the effort level $x_{21}^*$ of player 1 in group 2, and vice versa.

### 4.2.2. The case where $Z_1^* < 0$ and $Z_2^* < 0$

Using Lemmas 2, 4, and 6, we obtain Proposition 2.

**Proposition 2.** *Suppose that* $max\{v_{11}, \ v_{21}\} \leq \theta v_{1n_1} v_{2n_2}/|v_{1n_1} + v_{2n_2}|$ *or, equivalently,* $\theta \geq |v_{1n_1} + v_{2n_2}| \cdot max\{v_{11}, \ v_{21}\}/v_{1n_1} v_{2n_2}$. *Then, there exists a unique pure-strategy Nash equilibrium,* $(x_{11}^*, y_{11}^*, \ . \ . \ . \ , x_{1n_1}^*, y_{1n_1}^*, x_{21}^*, y_{21}^*, \ . \ . \ . \ , x_{2n_2}^*, y_{2n_2}^*)$, *such that, for* $i = 1, \ 2$, $y_{1n_1}^* = -v_{1n_1}^2 v_{2n_2}/(v_{1n_1} + v_{2n_2})^2$, $y_{2n_2}^* = -v_{1n_1} v_{2n_2}^2/(v_{1n_1} + v_{2n_2})^2$, $x_{ik}^* = 0$ *for* $k \in N_i$, *and* $y_{ik}^* = 0$ *for* $k \in N_i \setminus \{n_i\}$. *In this case, we have:* $X_i^* = 0$, $Y_i^* = y_{in_i}^*$, *and thus* $Z_i^* = -\theta y_{in_i}^* < 0$.

**Proof**. First, it is straightforward to check that $(x_{in_i}^*, y_{in_i}^*)$, for $i = 1, \ 2$, is the best response of player $n_i$ in group $i$ to the other players' equilibrium pairs of effort levels. Indeed, it is straightforward to check that $x_{in_i}^* = 0$ is her best response to the pair of $Z_{-in_i}^*$ and $Z_j^*$, for $j = 1, \ 2$ with $i \neq j$ (see the proof of part (*b*) of Lemma 2), and $y_{in_i}^*$ is her best response to the pair of $Z_{-in_i}^*$ and $Z_j^*$.

Next, it is also straightforward to check that $(x_{ih}^*, y_{ih}^*)$, for $i = 1, \ 2$ and $h \in N_i^- \setminus \{n_i\}$, is the best response of player $h$ to the other players' equilibrium pairs of effort levels (see the proof of part (*b*) of Lemma 2 and that of part (*a*) of Lemma 4).

Finally, we need to show that $(x_{ik}^*, y_{ik}^*)$, for $i = 1, \ 2$ and $k \in N_i^+$, is a best response of player $k$ to the other players' equilibrium pairs of effort levels. It is immediate from the proof of part (*a*) of Lemma 2 that $y_{ik}^* = 0$ is her best response to the pair of $Z_{-ik}^*$ and $Z_j^*$.

We now show that $x_{ik}^*$ is her best response to the pair of $Z_{-ik}^*$ and $Z_j^*$. Using Lemma 6, we know that $x_{ik}^* = 0$ is her best response to the pair of $Z_{-ik}^*$ and $Z_j^*$ if $v_{ik} \leq |Z_{-ik}^* + Z_j^*|$ $= \theta y_{1n_1}^* + \theta y_{2n_2}^* = -\theta v_{1n_1} v_{2n_2}/(v_{1n_1} + v_{2n_2}).$[10] This leads, under Assumption 1, to the fact that, for *every* player $k \in N_i^+$, $x_{ik}^*$ is her best response to the pair of $Z_{-ik}^*$ and $Z_j^*$ if $v_{i1} \le -\theta v_{1n_1} v_{2n_2}/(v_{1n_1} + v_{2n_2}).$

Therefore, for *both* $i = 1, 2$ and *every* player $k \in N_i^+$, $x_{ik}^*$ is her best response to the pair of $Z_{-ik}^*$ and $Z_j^*$ under the given condition that $\max\{v_{11}, v_{21}\} \le \theta v_{1n_1} v_{2n_2}/|v_{1n_1} + v_{2n_2}|.$   □

Note that $\theta v_{1n_1} v_{2n_2}/|v_{1n_1} + v_{2n_2}| < \theta|v_{in_i}|$ or, equivalently, $v_{1n_1} v_{2n_2}/|v_{1n_1} + v_{2n_2}| < |v_{in_i}|$ for $i = 1, 2$. Let $t = \max\{v_{11}, v_{21}\}$. Figure 2 illustrates, with the shaded area, the values of $|v_{1n_1}|$ and $|v_{2n_2}|$ which satisfy the condition that $t \le \theta v_{1n_1} v_{2n_2}/|v_{1n_1} + v_{2n_2}|$, given a value of $t$. Hence, the pure-strategy Nash equilibrium identified in Proposition 2 occurs at the values of $|v_{1n_1}|$ and $|v_{2n_2}|$ located in the shaded area of Figure 2.

If $t$ increases, then the curve representing the equation $\theta v_{1n_1} v_{2n_2}/|v_{1n_1} + v_{2n_2}| = t$ in Figure 2 shifts outward, reducing the shaded area. Suppose that $t = v_{11}$. In this case, if $v_{11}$ increases, *ceteris paribus*, then the best response of player 1 in group 1 to a pair of $Z_{-11}$ and $Z_2$ may change from 0 to $|Z_{-11}|$, depending on the given valuation profile (see Lemma 6). This implies that, if $v_{11}$ increases, the set of $(|v_{1n_1}|, |v_{2n_2}|)$ shrinks at which the pure-strategy Nash equilibrium identified in Proposition 2 occurs.

However, if the parameter $\theta$ increases, *ceteris paribus*, then the curve representing the equation $\theta v_{1n_1} v_{2n_2}/|v_{1n_1} + v_{2n_2}| = t$ in Figure 2 shifts inward, increasing the shaded area − that is, the set of $(|v_{1n_1}|, |v_{2n_2}|)$ expands at which the pure-strategy Nash equilibrium identified in Proposition 2 occurs. This makes sense because an increase in the parameter $\theta$ makes effort expended in sabotage activities more effective.

Proposition 2 says that, if $\max\{v_{11}, v_{21}\} \le \theta v_{1n_1} v_{2n_2}/|v_{1n_1} + v_{2n_2}|$, then there exists a pure-strategy Nash equilibrium with the following properties. First, there are only two active players, one from each group. The active player in group $i$ is player $n_i$ in that group − that is, the lowest-valuation player in the group. Second, every player in $N_i^-$ except player $n_i$ free rides on player $n_i$'s sabotage effort because she expects player $n_i$'s effort level to be large enough from her perspective (see Lemma 4). Third, every player in $N_i^+$ expends zero effort because her valuation $v_{ik}$ for the prize is far smaller than the "adjusted" valuation $\theta|v_{in_i}|$ of player $n_i$, the only active player, in her own group, and also because it is far smaller than the "adjusted" valuation $\theta|v_{jn_j}|$ of player $n_j$, the only active player, in the other group. To put it simply, she expends zero effort because she cannot defeat player $n_i$'s sabotage effort in her own group, and also because player $n_j$'s sabotage effort, in the other group, is strong enough from her perspective.

At the pure-strategy Nash equilibrium identified in Proposition 2, the two active players (or the two groups) compete to lose the prize rather than to win it − that is, they engage *only* in sabotage activities. We may simply say that the sabotage effort $y_{1n_1}^*$ of player $n_1$ in group 1 is the best response to the effort level $y_{2n_2}^*$ of player $n_2$ in group 2, and vice versa.

### 4.2.3. *Simple cases where the active players have the same valuations*

Consider first the case where $v_{11} = v_{21} = a$ and $|v_{1n_1}| = |v_{2n_2}| = b$. In this case, the pure-strategy Nash equilibrium identified in Proposition 1 occurs if $\theta \leq v_{11}v_{21}/(v_{11} + v_{21}) \cdot \max\{|v_{1n_1}|, |v_{2n_2}|\} = a/2b$. The pure-strategy Nash equilibrium identified in Proposition 2 occurs if $\theta \geq |v_{1n_1} + v_{2n_2}| \cdot \max\{v_{11}, v_{21}\}/v_{1n_1}v_{2n_2} = 2a/b$. Note that $2a/b > a/2b$. Note also that, given that $\theta = 1$, the pure-strategy Nash equilibrium identified in Proposition 1 occurs if $a \geq 2b$; the pure-strategy Nash equilibrium identified in Proposition 2 occurs if $b \geq 2a$.

Next, consider the case where $v_{11} = v_{21} = |v_{1n_1}| = |v_{2n_2}| = c$. In this case, the pure-strategy Nash equilibrium identified in Proposition 1 occurs if $\theta \leq 1/2$. The pure-strategy Nash equilibrium identified in Proposition 2 occurs if $\theta \geq 2$. Note that, given that $\theta = 1$, a pure-strategy Nash equilibrium does not occur.

### 4.3. *No pure-strategy Nash equilibrium of the game*

We end our analysis by identifying when a pure-strategy Nash equilibrium does not occur. The following remark comes from footnotes 8 and 10 and the fact that $v_{11}v_{21}/[(v_{11} + v_{21}) \cdot \max\{|v_{1n_1}|, \ |v_{2n_2}|\}] < |v_{1n_1} + v_{2n_2}| \cdot \max\{v_{11}, \ v_{21}\}/v_{1n_1}v_{2n_2}$ (see Appendix A).

**Remark 1.** *If $v_{11}v_{21}/[(v_{11} + v_{21}) \cdot max\{|v_{1n_1}|, \ |v_{2n_2}|\}] < \theta < |v_{1n_1} + v_{2n_2}| \cdot max\{v_{11}, \ v_{21}\}/ v_{1n_1}v_{2n_2}$, then a pure-strategy Nash equilibrium does not occur.*

Remark 1 and Propositions 1 and 2 state that the following sequence occurs as the parameter $\theta$ increases from zero: (i) At low values of $\theta$, a pure-strategy Nash equilibrium without sabotage activities occurs, (ii) then, no pure-strategy Nash equilibrium occurs, and (iii) at high values of $\theta$, a pure-strategy Nash equilibrium with sabotage activities occurs.

## 5. Conclusions

We have studied a contest between two groups over a group-specific public-good/bad prize. We have assumed the following. In each group, there are at least one player with a positive valuation for the prize and at least one player with a negative valuation. The players' valuations for the prizes are publicly known. Each player can exert two types of effort: one to help her own group win the prize, and one to hinder her own group from winning the prize. The groups (or players) play a noncooperative simultaneous-move game.

We have obtained two types of pure-strategy Nash equilibrium, depending on the parameter values: one *without* sabotage activities and one *with* sabotage activities. In the first type of equilibrium, only the highest-valuation player in each group expends positive effort to *help* her own group win the prize. However, in the second type, only the lowest-valuation player in each group expends positive effort to *hinder* her own group from winning the prize.

We have introduced a specific form of contest success function that determines each group's probability of winning the prize, taking into account players' sabotage activities. Can we modify it to study contests in which more than two groups compete over a group-specific public-good/bad prize? There may be some difficulties in doing so. For example, we may have a difficulty in defining each group's probability of winning the prize when all groups' *effective* effort levels are negative.

In this paper, we have assumed that the two groups are not vested with their initial probabilities of winning the prize. However, contests (over a group-specific public-good/bad prize) between groups with initial probabilities of winning are often observed. For example, in a competition for hosting an airport, some regions or areas may have an intrinsic advantage over others due to geological conditions. Accordingly, it would be interesting to consider a model in which the groups have initial probabilities of winning the prize.

Finally, it would be interesting to consider a model in which the contest success function for a group is a difference-form one or the selection rule of all-pay auctions which takes into account players' sabotage activities.

**Footnotes**

1.     In the theory of contests, a contest is formally defined as a situation in which players or groups of players compete by expending irreversible effort to win a prize. Examples include rent-seeking contests, environmental conflicts, elections, litigation, labor tournaments, patent-seeking contests, all-pay auctions, and sporting contests. Important work in the literature on the theory of contests includes Tullock (1980), Rosen (1986), Dixit (1987), Hillman and Riley (1989), Baik and Shogren (1992), Baye et al. (1993), Nitzan (1994), Moldovanu and Sela (2001), Szymanski (2003), Corchón (2007), Epstein and Nitzan (2007), Congleton et al. (2008), Siegel (2009), Konrad (2009), Vojnović (2015), and Beviá and Corchón (2024).

2.     Group $i$'s effective effort level $Z_i$ is positive, zero, or negative. If it is negative, for example, then effectiveness adjusted sabotage effort dominates constructive effort within group $i$. In this case, group $i$ as a whole seeks not to win the prize.

3.     The simplest logit-form contest success function that is extensively used in this literature takes the form: $p_i = s_i/S$ if $S > 0$, and $p_i = 1/n$ if $S = 0$, where $p_i$ represents the probability that player $i$ wins the prize, $s_i$ represents the effort level expended by player $i$, and $S \equiv \sum_{j=1}^{n} s_j$. See, for example, Tullock (1980), Hillman and Riley (1989), Katz et al. (1990), Epstein and Nitzan (2007), Epstein and Mealem (2009), Konrad (2009), Kolmar and Rommeswinkel (2013), Vojnović (2015), Balart et al. (2016), Dasgupta and Neogi (2018), Barbieri and Serena (2022), and Protopappas (2023).

4.     If $\theta|v_{ih}| = v_{i1}$, then we have: $\partial \pi_{ih}/\partial y_{ih} = 0$. However, since $\pi_{ih}$ is strictly convex in $y_{ih}$, as shown in Section 4.1.1, $y_{ih}^*$ is not a best response of player $h \in N_i^-$ to the list of the other players' equilibrium effort levels.

5.     Given $Z_j > 0$, group $i$'s optimal effective effort level $Z_i^b(k)$ for player $k \in N_i^+$ is defined as group $i$'s effective effort level that maximizes

$$v_{ik}p_i(Z_1, Z_2) \ - \ Z_i \ = \ v_{ik}Z_i/(Z_1 + Z_2) \ - \ Z_i.$$

That is, it is group *i*'s best response to $Z_j$ that is computed with $v_{ik}$. It is straightforward to check that, under Assumption 1, $Z_i^b(k) < Z_i^b(1)$ holds for $k \in N_i^+ \setminus \{1\}$.

6.     If $v_{ik} = \theta|v_{in_i}|$, then we have: $\partial \pi_{ik}/\partial x_{ik} = 0$. However, since $\pi_{ik}$ is strictly convex in $x_{ik}$, as shown in Section 4.1.2, $x_{ik}^*$ is not a best response of player $k \in N_i^+$ to the list of the other players' equilibrium effort levels.

7.     Recall from Lemma 1 that there exists no pure-strategy Nash equilibrium of the game such that $Z_i \geq 0$ and $Z_j \leq 0$ for $i, j = 1, 2$ with $i \neq j$.

8.     It follows from Lemma 5 that $Z_{-ih}^*/\theta > 0$ is her best response to the pair of $Z_{-ih}^*$ and $Z_j^*$ if $\theta|v_{ih}| \geq Z_{-ih}^* + Z_j^*$. This, together with part (*b*) of Lemma 3, leads to the following result: If $\max\{\theta|v_{1n_1}|,\ \theta|v_{2n_2}|\} > v_{11}v_{21}/(v_{11} + v_{21})$ or, equivalently, $\theta > v_{11}v_{21}/[(v_{11} + v_{21}) \cdot \max\{|v_{1n_1}|, |v_{2n_2}|\}]$, then there exists no pure-strategy Nash equilibrium in which sabotage activities do not occur.

9.     Another way to explain this is as follows. Consider the condition in Proposition 1 expressed as $\theta \leq v_{11}v_{21}/(v_{11} + v_{21})\ \max\{|v_{1n_1}|,\ |v_{2n_2}|\}$. Given a valuation profile, if the parameter $\theta$ increases, *ceteris paribus*, then the condition may not be satisfied, so that the pure-strategy Nash equilibrium identified in Proposition 1 may not occur.

10.     It follows from Lemma 6 that $|Z_{-ik}^*| > 0$ is her best response to the pair of $Z_{-ik}^*$ and $Z_j^*$ if $v_{ik} \leq |Z_{-ik}^* + Z_j^*|$. This, together with part (*b*) of Lemma 4, leads to the following result: If $\max\{v_{11},\ v_{21}\} > \theta v_{1n_1}v_{2n_2}/|v_{1n_1} + v_{2n_2}|$ or, equivalently, $\theta < |v_{1n_1} + v_{2n_2}| \cdot \max\{v_{11},\ v_{21}\}/v_{1n_1}v_{2n_2}$, then there exists no pure-strategy Nash equilibrium in which sabotage activities occur.

## Appendix A: Proof of the strict inequality in Section 4.3

We prove by contradiction that $v_{11}v_{21}/[(v_{11} + v_{21}) \cdot \max\{|v_{1n_1}|, |v_{2n_2}|\}] < |v_{1n_1} + v_{2n_2}| \cdot \max\{v_{11}, v_{21}\}/v_{1n_1}v_{2n_2}$. Suppose on the contrary that $v_{11}v_{21}/[(v_{11} + v_{21}) \cdot \max\{|v_{1n_1}|, |v_{2n_2}|\}] \geq |v_{1n_1} + v_{2n_2}| \cdot \max\{v_{11}, v_{21}\}/v_{1n_1}v_{2n_2}$. Then, using the fact that $v_{11}v_{21} = \max\{v_{11}, v_{21}\} \cdot \min\{v_{11}, v_{21}\}$ and the fact that $v_{1n_1}v_{2n_2} = \max\{|v_{1n_1}|, |v_{2n_2}|\} \cdot \min\{|v_{1n_1}|, |v_{2n_2}|\}$, we have

$$\max\{v_{11}, v_{21}\} \cdot \min\{v_{11}, v_{21}\}/[(v_{11} + v_{21}) \cdot \max\{|v_{1n_1}|, |v_{2n_2}|\}]$$
$$\geq |v_{1n_1} + v_{2n_2}| \cdot \max\{v_{11}, v_{21}\}/[\max\{|v_{1n_1}|, |v_{2n_2}|\} \cdot \min\{|v_{1n_1}|, |v_{2n_2}|\}].$$

This inequality is simplified to

$$\min\{v_{11}, v_{21}\}/(v_{11} + v_{21}) \geq |v_{1n_1} + v_{2n_2}|/\min\{|v_{1n_1}|, |v_{2n_2}|\}.$$

This leads to a contradiction, beacuse $\min\{v_{11}, v_{21}\}/(v_{11} + v_{21}) < 1$ and $|v_{1n_1} + v_{2n_2}|/\min\{|v_{1n_1}|, |v_{2n_2}|\} > 1$.

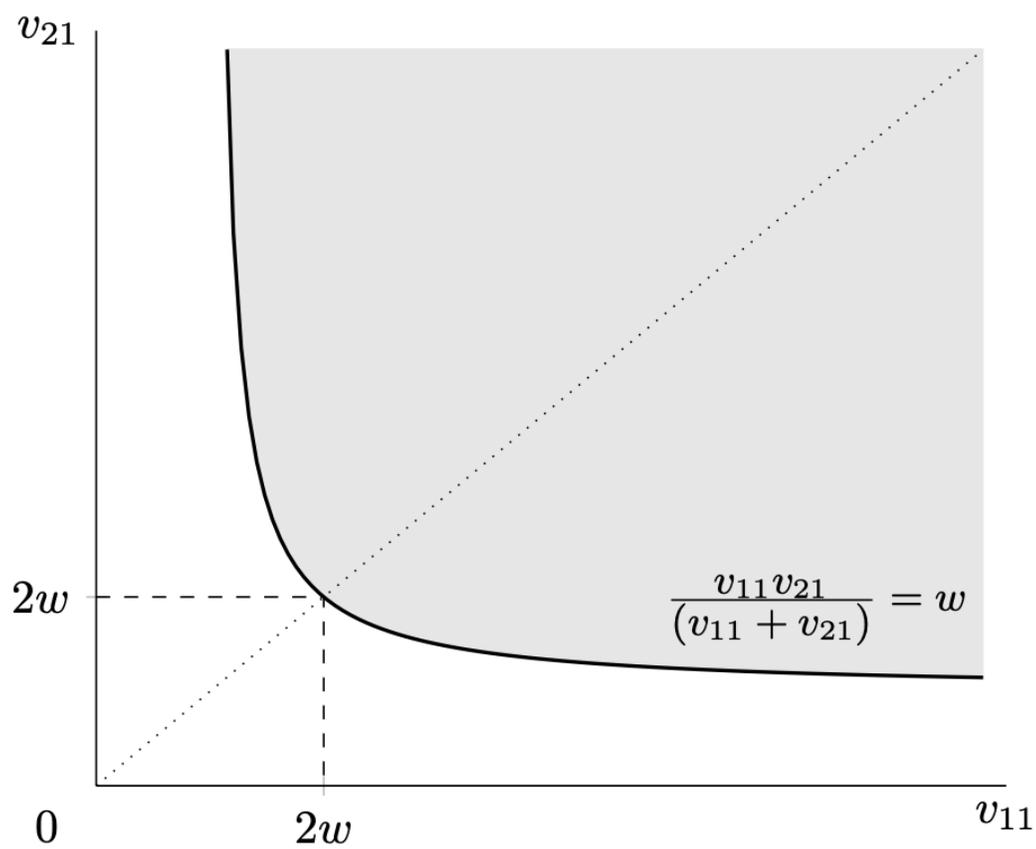

Figure 1. The Existence of a Pure-Strategy Nash Equilibrium with $Z_1^* > 0$ and $Z_2^* > 0$

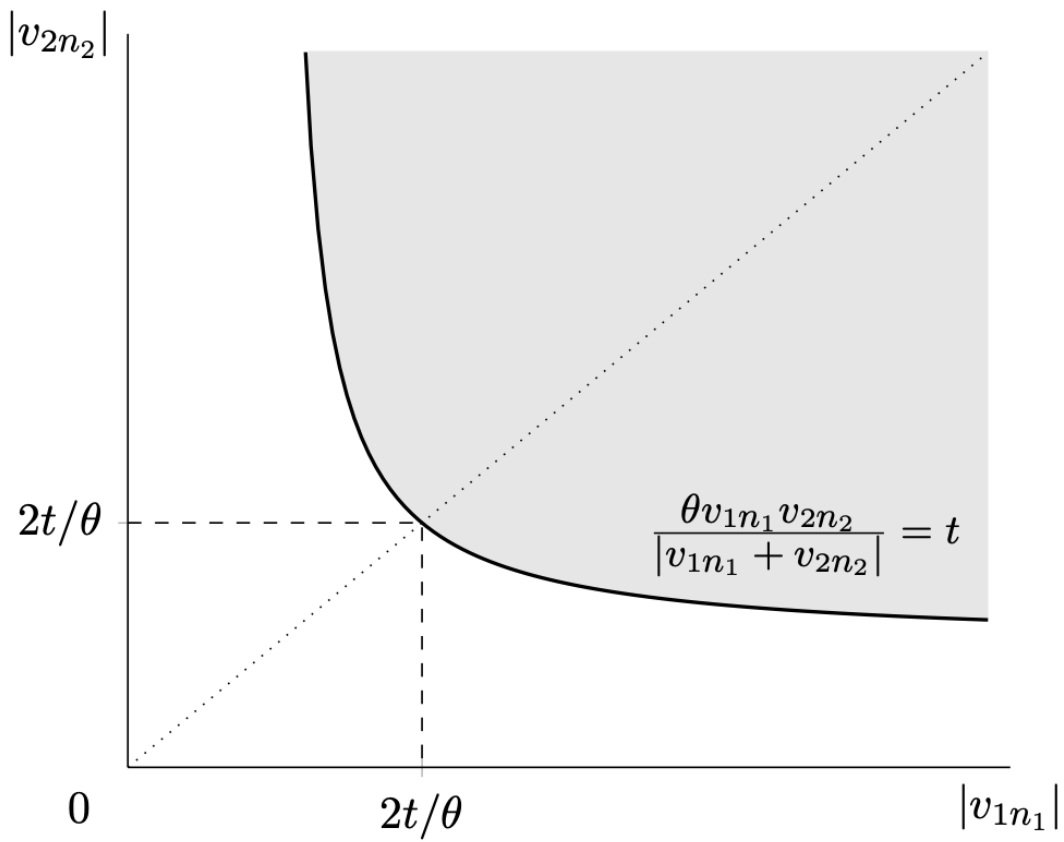

Figure 2. The Existence of a Pure-Strategy Nash Equilibrium with $Z_1^* < 0$ and $Z_2^* < 0$